\documentclass[10pt]{article}

\usepackage{graphics}
\usepackage{latexsym}
\usepackage{latexsym}
\usepackage{amssymb}
\usepackage{amsmath}
\usepackage{amsfonts}
\usepackage{epstopdf}
\usepackage[dvips]{graphicx}

\begin{document}

\newtheorem{remar}{\textbf{Remark}}[section]
\newtheorem{thar}{\textbf{Theorem}} [section]
\newtheorem{proar}{\textbf{Proposition}}[section]
\newtheorem{lemar}{\textbf{Lemma}}[section]
\newtheorem{corar}{\textbf{Corollary}}[section]

\author{$^{+}$Pierre~Gaillard,
$^{+}$ Universit\'{e} de Bourgogne, Dijon, France : \\
e-mail: Pierre.Gaillard@u-bourgogne.fr,\\
}

\title{\LARGE{\textbf{ Multi-parametric solutions to the NLS equation.} }  }


\maketitle

\begin{abstract}
The structure of the solutions to the one dimensional focusing
nonlinear Schr\"odinger equation (NLS) for the order $N$ in terms of
quasi rational functions is given here. We first give the proof that
the solutions can be expressed as a ratio of two wronskians of order
$2N$ and then two determinants by an exponential depending on $t$
with $2N-2$ parameters. It also is proved that for the order $N$,
the solutions can be written as the product of an exponential
depending on $t$ by a quotient of two polynomials of degree $N(N+1)$
in $x$ and $t$. The solutions depend on $2N-2$ parameters and give
when all these parameters are equal to $0$, the analogue of the
famous Peregrine breather $P_{N}$. It is fundamental to note that in
this representation at order $N$, all these solutions can be seen as
deformations with $2N-2$ parameters of the famous Peregrine breather
$P_{N}$. With this method, we already built Peregrine breathers
until order $N=10$, and their deformations depending on $2N-2$
parameters.
\end{abstract}

\section{Introduction}
The term of rogue wave was introduced in the scientific community by
Draper in $1964$ \cite{Draper}. The usual criteria for rogue waves
in the ocean, is that the vertical distance from trough to crest is
two or more times greater than the average wave height among one
third of the highest waves in a time series ($10$ to $30$ min). The
first rogue wave recorded by scientific measurement in North Sea was
made on the oil platform of Draupner in $1995$, located between
Norway and Scotland. Rogue waves in the ocean have led to many
marine catastrophes; it is one of the reasons why these rogue waves turn out
to be so important for the scientific community. It becomes a challenge to get a better understanding of
their mechanisms of formation. \\
The rogue waves phenomenon currently exceed the strict framework of
the study of ocean's waves and play a significant role in other
fields; in nonlinear optics \cite{Solli}, Bose-Einstein condensate
\cite{Bludov}, atmosphere  \cite{Stenflo} and even finance
\cite{Yan}.\\
Here, we consider the one dimensional focusing nonlinear
Schr\"odinger equation (NLS) to describe the phenomena of rogue
waves. The first results concerning the NLS equation date back the
works of Zakharov and Shabat in $1968$ who solved it using the
inverse scattering method \cite{Zakharov1, Zakharov2}. The case of
periodic and almost periodic algebro-geometric solutions to the
focusing NLS equation were first constructed in $1976$ by Its and
Kotlyarov \cite{its3, its1}. In $1977$ Kuznetsov found the first breather
type solution of the NLS equation \cite{Kuznetsov}; a simular result was given by Ma \cite{Ma} in $1979$. The first quasi rational solutions to NLS equation were constructed in $1983$ by
Peregrine \cite{Peregrine1}. In $1986$ Akhmediev, Eleonski and
Kulagin obtained the two-phase almost periodic solution to the NLS
equation and obtained the first higher order analogue of the
Peregrine breather \cite{Akhmediev8}. Other analogues of Peregrine breathers of order $3$ were constructed and initial data corresponding to orders $4$ and $5$ were described in a series of articles by Akhmediev
et al., in particular in \cite{Akhmediev1, Akhmediev15} using Darboux transformations. \\
Quite recently, many works about NLS equation have been published
using different methods. In $2010$, rational solutions to the NLS
equation were written as a quotient of two wronskians \cite{DGKM}.
In $2011$, the present author constructed in \cite{Gaillard10}
another representation of the solutions to the NLS equation in terms
of a ratio of two wronskians of even order $2N$ composed of
elementary functions using truncated Riemann theta functions depending on two parameters;
rational solutions were obtained when some parameter tended to $0$.
In $2012$, Guo, Ling and Liu found another representation of the
solutions as a ratio of two determinants \cite{Guo} using
generalized Darboux transform; a new approach was proposed by Ohta
and Yang in \cite{Ohta1} using Hirota bilinear method; finally, the
present author has obtained rational solutions in
terms of determinants which do not involve limits in \cite{Gaillard17} depending on two parameters. \\
With this extended method, we present multi-parametric families of quasi
rational solutions to the focusing NLS equation of order $N$ in terms of determinants
(determinants of order $2N$) dependent on $2N-2$ real parameters.
With this representation, at the same time
the well-known ring structure, but also the triangular shapes also
found by Ohta and Yang \cite{Ohta1}, Akhmediev et al.
\cite{Kedziora2} are given. \\
The aim of this paper is to prove the representation of the solutions to the focusing
NLS equation depending this time on $2N-2$ parameters; the proof presented in this paper with $2N-2$ parameters has been never published. This is the first task of the paper; then we deduce its particular degenerate representations in terms of a ratio of two determinants of order $2N$.
The second task of the paper is to give the proof of the structure of the solution at the order $N$ as the ratio of two polynomials of order $N(N+1)$ in $x$ and $t$ by an exponential depending on $t$. This representation makes possible to get all the possible patterns for the solutions to the NLS equation. It is important to stress that contrary to other methods, these solutions depending on $2N-2$ parameters give the Peregrine breather as parti\-cular case when all the parameters are equal to $0$ : for this reason, these solutions will be called $2N-2$ parameters deformations of the Peregrine of order $N$. \\
The paper is organized as follows. First of all, we express the
solutions of the NLS equation using Fredholm determinants from these
expressed in terms of truncated functions theta of Riemann first
obtained by Its, Rybin and Salle \cite{its1}; the representation
given in theorem $2.1$ is different from those given in \cite{its1}.
From that, we prove the representation of the solutions of the NLS equation in terms of wronskians depending on $2N-2$ parameters.
We deduce a degenerate representation of solutions to the NLS equation depending a priori on $2N-2$ parameters at the order $N$.\\
Then we prove a theorem which states the structure of the quasi-rational solutions to the NLS equation.
It was only conjectured in preceding works \cite{Gaillard10, Gaillard17}.
Families depending on $2N-2$ parameters for the $N$-th order as a ratio of two polynomials of $x$ and $t$
multiplied by an exponential depending on $t$ are obtained; it is proved that each of these polynomials have a degree equal to $N(N+1)$. \\

\section{Expression of solutions to the NLS equation in terms of wronskians}

\subsection{Solutions to the NLS equation in terms of $\theta$ functions}
For $r=1,3$, we define
\begin{eqnarray}
\label{thetai} 
\begin{array}{l} \theta_{r}(x,t) = \sum_{k \in
\{0;1\}^{2N} } \exp \left\{ \sum_{\mu
> \nu, \, \mu,\nu=1}^{2N}  \ln \left(
\frac{\gamma_{\nu}-\gamma_{\mu}}
{\gamma_{\nu}+\gamma_{\mu}}\right)^{2}  k_{\mu} k_{\nu} \right.
\\
\left. + \left( \sum_{\nu=1}^{2N} i \kappa_{\nu} x  - 2 \delta_{\nu}
t + x_{r,\nu} +  \sum_{\mu =1, \, \mu \neq \nu}^{2n} \ln \left|
\frac{\gamma_{\nu}+ \gamma_{\mu}} {\gamma_{\nu}- \gamma_{\mu}}
\right| + \pi i \epsilon_{\nu} +e_{\nu} \right) k_{\nu}  \right\} ,
\end{array}
\end{eqnarray}
In this formula, the symbol $\sum_{k \in
\{0;1\}^{2N} } $ denotes summation over all 2N-dimensional vectors
$k$ whose coordinates $k_{\nu}$ are either $0$ or $1$. \\
The terms $\kappa_{\nu},\delta_{\nu},\gamma_{\nu}$ and $x_{r,\nu}$
are functions of the parameters $\lambda_{\nu},\, 1\leq \nu \leq
2N$; they are defined by the formulas :
\begin{eqnarray}
\label{par3} \begin{array}{l} \kappa_{\nu} = 2
\sqrt{1-\lambda_{\nu}^{2}}, \hspace{0.5cm} \delta_{\nu} =
\kappa_{\nu} \lambda_{\nu}, \hspace{0.5cm}
\gamma_{\nu} = \sqrt{\frac{1-\lambda_{\nu}}{1+\lambda_{\nu}},};\\
x_{r,\nu} = (r-1) \ln  \frac{\gamma_{\nu} - i}{\gamma_{\nu} + i},
\quad r=1,3.
\end{array}
\end{eqnarray}
The parameters $-1<\lambda_{\nu}<1$, $\nu = 1,\ldots,2N$, are real
numbers such that
\begin{eqnarray}
\label{slambda}
\begin{array}{l}
-1 < \lambda_{N+1} < \lambda_{N+2} < \ldots < \lambda_{2N} < 0 <
\lambda_{N} < \lambda_{N-1} < \ldots < \lambda_{1} < 1\\
\lambda_{N+j} = -\lambda_{j}, \quad j=1,\ldots ,N.
\end{array}
\end{eqnarray}
The condition (\ref{slambda}) implies that
\begin{eqnarray}
\kappa_{j+N} =\kappa_{j}, \quad \delta_{j+N} =-\delta_{j+N}, \quad
\gamma_{j+N}=\gamma_{j}^{-1},\quad x_{r,j+N}= x_{r,j},  \quad
j=1,\ldots,N.
\end{eqnarray}
Complex numbers $e_{\nu}$ $1 \leq \nu \leq 2N$  are defined in the
following way :
\begin{eqnarray}
\label{ei}
\begin{array}{l}
e_{j}=ia_{j} - b_{j}, \quad e_{N+j}=ia_{j} + b_{j}, \quad 1 \leq j
\leq N, \quad a,b \in \bf{R}.
\end{array}
\end{eqnarray}
$\epsilon_{\nu} \in \{0;1\}$, $\varphi$, $\nu = 1 \ldots 2N$ are arbitrary real numbers.\\
With these notations, the solution of the NLS equation
\begin{eqnarray}
\label{NLS1} i v_{t}+ v_{xx} + 2|v|^{2}v=0,
\end{eqnarray}
can be expressed as (\cite{its1})
\begin{eqnarray}
\label{solNLS} v(x,t) = \frac{\theta_{3}(x,t)}{\theta_{1}(x,t)}
\exp(2it-i\varphi),
\end{eqnarray}

\subsection{From $\theta$ functions to Fredholm determinants}
To get Fredholm determinants, we have to express the functions
$\theta_{r}$ defined in (\ref{thetai}) in terms of subsets of
$[1,..,2N] $
\begin{eqnarray}
\label{thetaJ} \theta_{r}(x,t) = \sum_{J \subset \{1,..,2N\}  }
\prod_{\nu \in J}(-1)^{\epsilon_{\nu}}  \prod_{ \nu \in J,\, \mu
\notin J} \left| \frac{\gamma_{\nu}+\gamma_{\mu}}
{\gamma_{\nu}-\gamma_{\mu}}\right| \times \exp \left( \sum_{ \nu \in
J } i \kappa_{\nu} x  - 2 \delta_{\nu} t   + x_{r,\nu}\ + e_{\nu}
\right).
\end{eqnarray}
In (\ref{thetaJ}), the symbol $\sum_{J \subset \{1,..,2N\}  }$
denotes summation over all subsets $J$ of indices of the set $
\{1,..,2N\} $. \\
Let $I$ be the unit matrix and $C_{r}=(c_{jk})_{1\leq j,k\leq 2N}$
the matrix  defined by :
\begin{eqnarray}
\label{C} c_{\nu \mu}=   (-1)^{\epsilon_{\nu}} \frac{ \prod_{\eta
\neq \mu} \left| \gamma_{\nu}+\gamma_{\eta} \right|} {\prod_{\eta
\neq \nu} \left|\gamma_{\nu}-\gamma_{\eta}\right| }\exp(   i
\kappa_{\nu} x -2 \delta_{\nu}t + x_{r,\nu} + e_{\nu}),
\end{eqnarray}
\begin{eqnarray}
\label{eps} \epsilon_{j} = j \quad 1\leq j \leq N, \quad
\epsilon_{j} = N+j, \quad N+1\leq j \leq 2N.
\end{eqnarray}
Then $\det(I+C_{r})$ has the following form
\begin{eqnarray}
\label{fre} \det(I+C_{r}) = \sum_{J \subset \{1,...,2N\}} \prod_{\nu
\in J} (-1)^{\epsilon_{\nu}} \prod_{\nu \in J\, \mu \notin J} \left|
\frac{ \gamma_{\nu} + \gamma_{\mu} }{ \gamma_{\nu} - \gamma_{\mu} }
\right| \exp(   i \kappa_{\nu} x  -2 \delta_{\nu}t + x_{r,\nu} +
e_{\nu}).
\end{eqnarray}
Comparing this last expression (\ref{fre}) with the formula
(\ref{thetaJ}) at the beginning of this section, we have clearly the
identity
\begin{eqnarray}
\label{fre-the} \theta_{r} = \det (I+C_{r}).
\end{eqnarray}
We can give another representation of the solutions to NLS equation.
To do this, let's consider the matrix $D_{r}=(d_{jk})_{1\leq j,k\leq
2N}$ defined by :
\begin{eqnarray}
\label{defD} d_{\nu \mu}=   (-1)^{\epsilon_{\nu}} \prod_{\eta \neq
\mu}
\left|\frac{\gamma_{\eta}+\gamma_{\nu}}{\gamma_{\eta}-\gamma_{\mu}}\right|
\exp(   i \kappa_{\nu} x  -2 \delta_{\nu}t + x_{r,\nu} + e_{\nu}).
\end{eqnarray}
We have the equality $\det (I+D_{r}) = \det (I+C_{r})$, and so the
solution of NLS equation takes the form
\begin{eqnarray}
\label{ivt2} v(x,t) = \frac{\det (I+D_{3}(x,t))}{\det
(I+D_{1}(x,t))} \exp(2it-i\varphi).
\end{eqnarray}
\begin{thar}
The function $v$ defined by
\begin{eqnarray}
\label{solF} v(x,t) = \frac{\det (I+D_{3}(x,t))}{\det
(I+D_{1}(x,t))} \exp(2it-i\varphi).
\end{eqnarray}
is a solution of the focusing NLS equation with the matrix
$D_{r}=(d_{jk})_{1\leq j,k\leq 2N}$ defined by
\begin{eqnarray}
d_{\nu \mu}=   (-1)^{\epsilon_{\nu}} \prod_{\eta \neq \mu}
\left|\frac{\gamma_{\eta}+\gamma_{\nu}}{\gamma_{\eta}-\gamma_{\mu}}\right|
\exp(   i \kappa_{\nu} x  -2 \delta_{\nu}t + x_{r,\nu} + e_{\nu}).
\nonumber
\end{eqnarray}
where $\kappa_{\nu}$, $\delta_{\nu}$, $x_{r,\nu}$, $\gamma_{\nu}$,
$e_{\nu}$  being defined in(\ref{par3}), (\ref{slambda}) and
(\ref{ei}).
\end{thar}

\subsection{From Fredholm determinants to wronskians}
We want to express solutions to NLS equation in terms of wronskian
determinants. For this, we need the following notations :
\begin{eqnarray}
\begin{array}{l}
\label{phiPG} \phi_{r,\nu}= \sin   \Theta_{r,\nu} , \quad 1\leq \nu
\leq N,  \quad \phi_{r,\nu} = \cos  \Theta_{r,\nu} , \quad N+1\leq
\nu \leq 2N, \quad r=1, 3,
\end{array}
\end{eqnarray}
with the arguments
\begin{eqnarray}
\begin{array}{l}
\Theta_{r,\nu} =   \kappa_{\nu} x /2 +i \delta_{\nu} t -i
x_{r,\nu}/2 + \gamma_{\nu} y - i e_{\nu}/2 , \quad 1 \leq \nu \leq
2N.
\end{array}
\end{eqnarray}
We denote $W_{r} (y)$ the wronskian of the functions $
\phi_{r,1},\ldots,\phi_{r,2N} $ defined by \\
\begin{eqnarray}
\label{W}
W_{r}(y) = \det
[(\partial^{\mu-1}_{y}\phi_{r,\nu})_{\nu,\, \mu \in [1,\ldots,2N]}
].
\end{eqnarray}
We consider the matrix $D_{r}=(d_{\nu \mu})_{\nu,\,
\mu \in [1,\ldots,2N]}$ defined in (\ref{defD}). Then we have the
following statement
\begin{thar}
\begin{eqnarray}
\label{FW} \det (I+ D_{r}) = k_{r} (0) \times
W_{r}(\phi_{r,1},\ldots,\phi_{r,2N}) (0),
\end{eqnarray}
where
$$
k_{r}(y) = \frac{2^{2N}  \exp (i \sum_{\nu=1}^{2N} \Theta_{r,\nu} )
}{ \prod_{\nu=2}^{2N} \prod_{\mu=1}^{\nu-1} ( \gamma_{\nu} -
\gamma_{\mu} ) }.
$$
\end{thar}
\noindent \textbf{Proof : } We start to remove the factor
$(2i)^{-1}e^{i \Theta_{r,\nu}}$ in each row $\nu$ in the wronskian
$W_{r}(y)$ for $1\leq \nu \leq 2N
$.\\
Then
\begin{eqnarray}
\label{1W} W_{r} = \prod_{\nu=1}^{2N} e^{i \Theta_{r,\nu}} (2i)^{-N}
(2)^{-N} \times \tilde{W}_{r},
\end{eqnarray}
with
$$
\tilde{W}_{r} = \left|\begin{array}{cccc}
(1-e^{-2 i \Theta_{r,1} }) & i  \gamma_{1}(1+e^{-2 i \Theta_{r,1} }) & \ldots & (i \gamma_{1})^{2N-1}(1+(-1)^{2N} e^{-2 i \Theta_{r,1} })\\
(1-e^{-2 i \Theta_{r,2} }) & i  \gamma_{2}(1+e^{-2 i \Theta_{r,2} }) & \ldots & (i  \gamma_{2})^{2N-1}(1+(-1)^{2N} e^{-2 i \Theta_{r,2} })\\
\vdots & \vdots & \vdots & \vdots \\
(1-e^{-2 i \theta_{r,2N} }) & i  \gamma_{2N}(1+e^{-2 i \Theta_{r,2N}
}) & \ldots & (i  \gamma_{2N})^{2N-1}(1+(-1)^{2n} e^{-2 i
\Theta_{r,2N} })
\end{array}
\right|
$$
The determinant $\tilde{W}_{r}$ can be written as
$$
\tilde{W}_{r} = \det (\alpha_{jk} e_{j} + \beta_{jk}),
$$
where $\alpha_{jk} = (-1)^{k} (i \gamma_{j})^{k-1}$, $e_{j} = e^{-2
i \Theta_{r,j} }$,
and $\beta_{jk} = (i \gamma_{j})^{k-1}$, $1\leq j \leq N$, $1\leq k \leq 2N$,\\
$\alpha_{jk} = (-1)^{k-1} (i \gamma_{j})^{k-1}$, $e_{j} = e^{-2 i
\Theta_{r,j} }$,
and $\beta_{jk} = (i \gamma_{j})^{k-1}$, $N+1\leq j \leq 2N$, $1\leq k \leq 2N$.\\
We want to calculate $\tilde{W}_{r}$. To do this, we use the
following Lemma
\begin{lemar}
Let $A = (a_{ij})_{i,\, j \in [1,...,N]}$, $B = (b_{ij})_{i,\, j \in [1,...,N]}$, \\
$(H_{ij})_{i,\, j \in [1,...,N]}$, the matrix formed by replacing in
$A$ the jth row of $A$ by the ith row of $B$  Then
\begin{eqnarray}
\label{lem} \det (a_{ij} x_{i} + b_{ij}) = \det (a_{ij}) \times \det
(\delta_{ij} x_{i} + \frac{\det(H_{ij})}{\det ( a_{ij} ) } )
\end{eqnarray}
\end{lemar}
\noindent \textbf{Proof :} We use the classical notations :
$\tilde{A} = (\tilde{a}_{ji})_{i,\, j \in [1,...,N]}$ the transposed
matrix in cofactors of $A$. We have the well known formula
$A \times \tilde{A} = \det{A} \times I$.\\
So it is clear that $\det (\tilde{A}) = (\det(A))^{N-1}$.\\
The general term of the product $(c_{ij})_{i,j \in [1,..,N]} =
{(a_{ij} x_{i} + b_{ij} )}_{i,j \in [1,..,N]}
\times {(\tilde{a}_{ji}})_{i,j \in [1,..,N]}$ can be written as  \\
$c_{ij} = \sum_{s=1}^{N}(a_{is} x_{i} + b_{is} ) \times \tilde{a}_{js}  \\
= x_{i} \sum_{s=1}^{n} a_{is} \tilde{a}_{js}  +  \sum_{s=1}^{n} b_{is} \tilde{a}_{js} \\
= \delta_{ij} \det (A) x_{i} + \det (H_{ij}) $.\\
We get  \\
$\det (c_{ij}) =  \det (a_{ij} x_{i} + b_{ij} )
\times  (\det(A))^{N-1}=  (\det(A))^{N} \times \det (\delta_{ij} x_{i} + \frac{\det(H_{ij})}{\det(A)}  )$. \\
Thus $\det (a_{ij} x_{i} + b_{ij} )
=  \det(A) \times \det (\delta_{ij} x_{i} + \frac{\det(H_{ij})}{\det(A)}  )$. \\
$\Box$.\\
We denote $U = (\alpha_{ij})_{i,\, j \in [1,...,2N]}$, $V =
(\beta_{ij})_{i,\, j \in [1,...,2N]}$. \\
By applying the previous lemma, one obtains :
\begin{eqnarray}
\label{wr1}
\begin{array}{l}
\tilde{W}_{r} = \det (\alpha_{ij} e_{i} + \beta_{ij})  \\
= \det (\alpha_{ij}) \times \det (\delta_{ij} e_{i} +
\frac{\det(H_{ij})}{\det ( \alpha_{ij} ) } )= \det (U) \times \det
(\delta_{ij} e_{i} + \frac{\det(H_{ij})}{\det ( U ) } ),
\end{array}
\end{eqnarray} where
$(H_{ij})_{i,\, j \in [1,...,N]}$ is the matrix formed by replacing
in $U$ the jth row of $U$ by the ith row of $V$
defined previously. \\
The determinant of $U$ of Vandermonde type is clearly equal to
\begin{eqnarray}
\label{U} \det (U) =  i^{ N(2N-1) } \prod_{2N\geq l > m \geq 1} (
\gamma_{l} -  \gamma_{m}).
\end{eqnarray}
To calculate determinant $\tilde{W}_{r}$, we must compute now
$\det (H_{ij})$. To do that, two cases must be studied : \\
1. For $1 \leq  j  \leq N$. The matrix $H_{ij}$ is clearly of the
VanderMonde type where the $j$-th row of $U$ in $U$ is replaced by
the $i$-th row of $V$. Clearly, we have :
\begin{eqnarray}
\label{H}
\begin{array}{l}  \det (H_{ij}) = (-1) ^{
N(2N+1) + N-1 } (i) ^{ N(2N-1)  } \times  M,
\end{array}
\end{eqnarray}
where $M = M(m_{1},\ldots, m_{2N})$ is the Vandermonde determinant
defined by $m_{k}=\gamma_{k}$ for $k \neq j$ and
$m_{j}=-\gamma_{i}$. Thus we have :
\begin{eqnarray}
\begin{array}{l}
\det (H_{ij}) = -(i) ^{ N(2N-1)} \times  \prod_{2N \geq l
> k \geq 1,} \, ( m_{l} - m_{k} )\\
\\
= -(i) ^{ N(2N-1)  } \times \prod_{ 2N \geq l
> m \geq 1 , \, l\neq j, \,
m \neq j} (  \gamma_{l} -  \gamma_{m})  \times \prod_{l < j} (
-\gamma_{i} - \gamma_{l}) \times \prod_{l > j} ( \gamma_{l} +
\gamma_{i}), \\
\\
= (-1)^{j} (i) ^{ N(2N-1)  } \times \prod_{ 2N \geq l
> m \geq 1 , \, l\neq j, \,
m \neq j} (  \gamma_{l} -  \gamma_{m})  \times \prod_{l \neq j } (
\gamma_{l} + \gamma_{i}).
\end{array}
\end{eqnarray}
To evaluate $\tilde{W}_{r}$, we must simplify the quotient $q_{ij} :
= \frac{\det (H_{ij})}{\det (U)}$ :
\begin{eqnarray}
\begin{array}{l}
 q_{ij} = \frac{ (-1)^{j} (i) ^{ N(2N-1)  } \times \prod_{ 2N \geq l
> m \geq 1 , \, l\neq j, \,
m \neq j} (  \gamma_{l} -  \gamma_{m})  \times \prod_{l \neq j } (
\gamma_{l} + \gamma_{i}) } { i^{ N(2N-1) } \prod_{2N\geq l > m \geq
1} (
\gamma_{l} -  \gamma_{m}) } \\
\\
= \frac{ (-1)^{j}  \prod_{l \neq j } ( \gamma_{l} + \gamma_{i}) } {
 \prod_{ l <j } ( \gamma_{j} -
\gamma_{l})  \prod_{ l > j } ( \gamma_{l} - \gamma_{j})}  = \frac{
(-1)^{j}  \prod_{l \neq j } ( \gamma_{l} + \gamma_{i}) } {
(-1)^{j-1} \prod_{l \neq j } ( \gamma_{l} - \gamma_{j}) }= - \frac{
\prod_{l \neq j } ( \gamma_{l} + \gamma_{i}) } {  \prod_{l \neq j }
( \gamma_{l} - \gamma_{j}) }.
\end{array}
\end{eqnarray}
We can replace $q_{ij}$ by $r_{ij}$ defined by $- \frac{ \prod_{l
\neq j } ( \gamma_{l} + \gamma_{i}) } {  \prod_{l \neq i } (
\gamma_{l} - \gamma_{i}) }$, because $\det (\delta_{ij} x_{i} +
\frac{\det(q_{ij})}{\det(A)}  ) = \det (\delta_{ij} x_{i} +
\frac{\det(r_{ij})}{\det(A)}  )$ (similar matrices). \\
We express $r_{ij}$ in terms of absolute value; as $j \in[1;N]$ and
$0< \gamma_{1} < \ldots < \gamma_{N}<1 <\gamma_{2N} < \ldots <
\gamma_{N+1}$, we have :
\begin{eqnarray}
\begin{array}{l}
\prod_{l \neq i } ( \gamma_{l} - \gamma_{i}) = (-1)^{i-1} \prod_{l
\neq i } \left| \gamma_{l} - \gamma_{i} \right|,\quad \prod_{l \neq
j } ( \gamma_{l} + \gamma_{i}) = \prod_{l \neq j } \left| \gamma_{l}
+ \gamma_{i} \right|.
\end{array}
\end{eqnarray}
So the term $r_{ij}$ can be written as
\begin{eqnarray}
\begin{array}{l}
r_{ij} = (-1)^{i}  \frac{  \prod_{l \neq j } \left| \gamma_{l} +
\gamma_{i} \right| } { \prod_{l \neq i }\left| \gamma_{l} -
\gamma_{i} \right| }  = (-1)^{\epsilon(i)} \frac{  \prod_{l \neq j }
\left| \gamma_{l} + \gamma_{i} \right| } { \prod_{l \neq i }\left|
\gamma_{l} - \gamma_{i} \right| } = c_{ij}e^{-2i\Theta_{r,i}(0)},
\end{array}
\end{eqnarray}
with respect to the notations  given in (\ref{eps}) and
(\ref{defD}). \\

2. The same estimations for $ N+1 \leq j \leq 2N $ are made; $ \det
H_{ij}$ is first
\begin{eqnarray}
\label{H3}
\begin{array}{l}
\det (H_{ij}) = (-1) ^{ N(2N+1) + N-1 } (i) ^{ N(2N-1)  } \times  M,
\end{array}
\end{eqnarray}
with $M = M(m_{1},\ldots, m_{2N})$ the Vandermonde determinant
defined by $m_{k}=\gamma_{k}$ for $k \neq j$ and
$m_{j}=-\gamma_{i}$. Thus we have :
\begin{eqnarray}
\begin{array}{l}
\det (H_{ij}) = (i) ^{ N(2N-1)} \times  \prod_{2N \geq l
> k \geq 1,} \, ( m_{l} - m_{k} )\\
\\
= (i) ^{ N(2N-1)  } \times \prod_{ 2N \geq l
> m \geq 1 , \, l\neq j, \,
m \neq j} (  \gamma_{l} -  \gamma_{m})  \times \prod_{l < j} (
-\gamma_{i} - \gamma_{l}) \times \prod_{l > j} ( \gamma_{l} +
\gamma_{i}), \\
\\
= (-1)^{j-1} (i) ^{ N(2N-1)  } \times \prod_{ 2N \geq l
> m \geq 1 , \, l\neq j, \,
m \neq j} (  \gamma_{l} -  \gamma_{m})  \times \prod_{l \neq j } (
\gamma_{l} + \gamma_{i}).
\end{array}
\end{eqnarray}
The quotient $q_{ij} : = \frac{\det (H_{ij})}{\det (U)}$ equals :
\begin{eqnarray}
\begin{array}{l}
 q_{ij} = \frac{ (-1)^{j-1} (i) ^{ N(2N-1)  } \times \prod_{ 2N \geq l
> m \geq 1 , \, l\neq j, \,
m \neq j} (  \gamma_{l} -  \gamma_{m})  \times \prod_{l \neq j } (
\gamma_{l} + \gamma_{i}) } { i^{ N(2N-1) } \prod_{2N\geq l > m \geq
1} (
\gamma_{l} -  \gamma_{m}) } \\
\\
= \frac{ (-1)^{j-1}  \prod_{l \neq j } ( \gamma_{l} + \gamma_{i}) }
{
 \prod_{ l <j } ( \gamma_{j} -
\gamma_{l})  \prod_{ l > j } ( \gamma_{l} - \gamma_{j})}  = \frac{
(-1)^{j-1}  \prod_{l \neq j } ( \gamma_{l} + \gamma_{i}) } {
(-1)^{j-1} \prod_{l \neq j } ( \gamma_{l} - \gamma_{j}) }=  \frac{
\prod_{l \neq j } ( \gamma_{l} + \gamma_{i}) } {  \prod_{l \neq j }
( \gamma_{l} - \gamma_{j}) }.
\end{array}
\end{eqnarray}
We replace $q_{ij}$ by $r_{ij}$ defined by $ \frac{ \prod_{l \neq j
} ( \gamma_{l} + \gamma_{i}) } {  \prod_{l \neq i } ( \gamma_{l} -
\gamma_{i}) }$, for the same reason as previously exposed. \\
$r_{ij}$ is expressed in terms of absolute value; as $j \in[N+1;2N]$
and $0< \gamma_{1} < \ldots < \gamma_{N}<1 <\gamma_{2N} < \ldots <
\gamma_{N+1}$, we have :
\begin{eqnarray}
\begin{array}{l}
\prod_{l \neq i } ( \gamma_{l} - \gamma_{i}) = (-1)^{2N-i+N}
\prod_{l \neq i } \left| \gamma_{l} - \gamma_{i} \right|,\quad
\prod_{l \neq j } ( \gamma_{l} + \gamma_{i}) = \prod_{l \neq j }
\left| \gamma_{l} + \gamma_{i} \right|.
\end{array}
\end{eqnarray}
So the term $r_{ij}$ can be written as
\begin{eqnarray}
\begin{array}{l}
r_{ij} = (-1)^{N+i}  \frac{  \prod_{l \neq j } \left| \gamma_{l} +
\gamma_{i} \right| } { \prod_{l \neq i }\left| \gamma_{l} -
\gamma_{i} \right| }  = (-1)^{\epsilon(i)} \frac{  \prod_{l \neq j }
\left| \gamma_{l} + \gamma_{i} \right| } { \prod_{l \neq i }\left|
\gamma_{l} - \gamma_{i} \right| } = c_{ij}e^{-2i\Theta_{r,i}(0)},
\end{array}
\end{eqnarray}
with respect to the notations  given in (\ref{eps}) and
(\ref{defD}).\\
Replacing $e_{i}$ by $e^{-2i\Theta_{r,i}}$, $\det \tilde{W}_{r} $
can be expressed as
\begin{eqnarray}
\begin{array}{l}
\label{RFTH} \det \tilde{W}_{r} = \det (U) \times \det (\delta_{ij}
e_{i} + \frac{\det(H_{ij})}{\det ( U ) } ) = \det (U) \times \det
(\delta_{ij}
e_{i} + r_{ij} ) \\
= \det (U) \prod_{i=1}^{2N} e^{-2 i \Theta_{i}} \det ( \delta_{ij} +
 (-1) ^{  \epsilon(i) } \prod_{ l\neq i } \left| \frac{ \gamma_{l} +
\gamma_{i} } { \gamma_{l} - \gamma_{i}} \right|  e^{2 i
\Theta_{r,i}} ).
\end{array}
\end{eqnarray}
We estimate the two members of the last relation (\ref{RFTH}) in
$y=0$, and using (\ref{U}) we obtain the following result
\begin{eqnarray}
\begin{array}{l}
\det \tilde{W}_{r}(0) = i^{ N(2N-1) } \prod_{2N \geq l > m \geq 1} (
\gamma_{l} - \gamma_{m})  \prod_{i=1}^{2N} e^{-2 i \Theta_{r,i}(0)}
\\
\times \det ( \delta_{ij} + (-1) ^{  \epsilon(i) } \prod_{ l\neq i }
\left| \frac{ \gamma_{l} + \gamma_{i} } { \gamma_{l} - \gamma_{i}}
\right| e^{2 i \Theta_{r,i}(0)} ) \\
=  i^{ N(2N-1) } \prod_{j=2}^{2N} \prod_{i=1}^{j-1} ( \gamma_{j} -
\gamma_{i} )    e^{-2i \sum_{i=1}^{2N} \Theta_{r,i}(0)}  \det ( \delta_{ij} + c_{ij} ) \\
\\
=  i^{ N(2N-1) }  \prod_{j=2}^{2N} \prod_{i=1}^{j-1} ( \gamma_{j} -
\gamma_{i} )    e^{-2i \sum_{i=1}^{2N} \Theta_{r,i}(0)}  \det ( I+C_{r} )\\
\\
=  i^{ N(2N-1) } \prod_{j=2}^{2N} \prod_{i=1}^{j-1} ( \gamma_{j} -
\gamma_{i} )  e^{-2i \sum_{i=1}^{2N} \Theta_{r,i}(0)} \det ( I+D_{r}
).
\end{array}
\end{eqnarray}
Therefore, the wronskian $W_{r}$ given by (\ref{1W} ) can be written
as
\begin{eqnarray}
\begin{array}{l}
W_{r}(\phi_{r,1}, \ldots,\phi_{r,2N}) (0) =  \prod_{j=1}^{2N} e^{ i
\Theta_{r,j}(0) } (2)^{-2N} (i)^{-N} \times \tilde{W}_{}r\\
\\
=\prod_{j=1}^{2N} e^{ i \Theta_{r,j}(0) } (2)^{-2N} (i)^{-N} i^{
N(2N-1) } \prod_{j=2}^{2N} \prod_{i=1}^{j-1} ( \gamma_{j} -
\gamma_{i} )  e^{-2i \sum_{i=1}^{2N} \Theta_{r,i}(0)} \det ( I+D_{r}
)\\
\\
=(2)^{-2N} \prod_{j=2}^{2N} \prod_{i=1}^{j-1} ( \gamma_{j} -
\gamma_{i} )  e^{-i \sum_{i=1}^{2N} \Theta_{r,i}(0)} \det ( I+D_{r}
).
\end{array}
\end{eqnarray}
As a consequence
\begin{eqnarray}
\det(I+D_{r}) =  k_{r}(0) W_{r}(\phi_{1}, \ldots,\phi_{2N})(0).
\end{eqnarray}
$\Box$\\

\subsection{Wronskian representation of solutions to the NLS equation}
From the initial formulation (\ref{solF}) we have
$$
v(x,t) = \frac{\det (I+D_{3}(x,t))}{\det (I+D_{1}(x,t))}
\exp(2it-i\varphi).
$$
Using (\ref{FW}), the following relation between Fredholm
determinants and wronskians is obtained
$$
\det (I+ D_{3}) = k_{3} (0) \times
W_{3}(\phi_{r,1},\ldots,\phi_{r,2N}) (0)
$$
and
$$
\det (I+ D_{3}) = k_{3} (0) \times
W_{3}(\phi_{r,1},\ldots,\phi_{r,2N}) (0).
$$
As $\Theta_{3,j}(0)$ contains $N$ terms $x_{3,j}$ $1 \leq j \leq N$
and $N$ terms $-x_{3,j}$ $1 \leq j \leq N$, we have the equality
$k_{3}(0) = k_{1}(0)$, and we get the following result :
\begin{thar}
The function $v$ defined by
$$
v(x,t) =  \frac{W_{3}(\phi_{3,1}, \ldots,\phi_{3,2N}) (0) } {
W_{1}(\phi_{1,1}, \ldots,\phi_{1,2N}) (0) } \exp (2it -i \varphi).
$$
is a solution of the focusing NLS equation depending on two real
parameters $a$ and $b$ with $\phi^{r}_{\nu}$ defined in
(\ref{phiPG})
\begin{eqnarray}
\begin{array}{l}
\phi_{r,\nu}= \sin (\kappa_{\nu} x /2 +i \delta_{\nu} t -i
x_{r,\nu}/2 + \gamma_{\nu} y - i e_{\nu}/2 ) , \quad 1\leq \nu \leq
N, \nonumber \\
\phi_{r,\nu} = \cos ( \kappa_{\nu} x /2 +i \delta_{\nu} t -i
x_{r,\nu}/2 + \gamma_{\nu} y - i e_{\nu}/2  ) , \quad N+1\leq \nu
\leq 2N, \quad r=1, 3,
\end{array}
\end{eqnarray}
$\kappa_{\nu}$, $\delta_{\nu}$, $x_{r,\nu}$, $\gamma_{\nu}$,
$e_{\nu}$  being  defined in(\ref{par3}), (\ref{slambda}) and
(\ref{ei}).
\end{thar}


\section{Families of multi-parametric solutions to the NLS equation in
terms of a ratio of two determinants}
Solutions to the NLS equation as a quotient of two determinants are constructed. \\
Similar functions defined in a preceding work \cite{Gaillard17} are
used, but modified as explained in the following.
The following notations are needed : \\
$$
X_{\nu} = \kappa_{\nu} x/2 +i \delta_{\nu}t - i x_{3,\nu}/2  - i
e_{\nu}/2,
$$
$$
Y_{\nu} = \kappa_{\nu} x/2 +i \delta_{\nu}t  - i x_{1,\nu}/2 - i
e_{\nu}/2,
$$
for $1 \leq \nu \leq 2N$, with $\kappa_{\nu}$, $\delta_{\nu}$,
$x_{r,\nu}$ defined in (\ref{par3}). \\
Parameters $e_{\nu}$ are defined by (\ref{ei}). \\
Here, is the crucial point : we choose the parameters $a_{j}$ and
$b_{j}$ in the form
\begin{eqnarray}
\label{ej2} a_{j} = \sum_{k=1}^{N-1} \tilde{a_{k}} j^{2k+1}
\epsilon^{2k+1}, \quad b_{j} =  \sum_{k=1}^{N-1} \tilde{b_{k}}
j^{2k+1} \epsilon^{2k+1}, \quad 1 \leq j \leq N.
\end{eqnarray}
Below the following functions are used : \\
\begin{eqnarray}
\label{fonc1}
\begin{array}{l} \varphi_{4j+1,k} = \gamma_{k}^{4j-1}
\sin X_{k}, \quad \varphi_{4j+2,k} =
\gamma_{k}^{4j} \cos X_{k},  \\
\varphi_{4j+3,k} = - \gamma_{k}^{4j+1} \sin X_{k}, \quad
\varphi_{4j+4,k} = - \gamma_{k}^{4j+2} \cos X_{k},
\end{array}
\end{eqnarray}
for $1 \leq k \leq N$, and
\begin{eqnarray}
\label{fonc2}
\begin{array}{l} \varphi_{4j+1,N+k} =
\gamma_{k}^{2N-4j-2} \cos X_{N+k}, \quad \varphi_{4j+2,N+k} =-
\gamma_{k}^{2N-4j-3} \sin X_{N+k},\\
\varphi_{4j+3,N+k} = - \gamma_{k}^{2N-4j-4} \cos X_{N+k}, \quad
\varphi_{4j+4,N+k} = \gamma_{k}^{2N-4j-5} \sin X_{N+k},
\end{array}
\end{eqnarray}
for $ 1 \leq k \leq N$.\\
We define the functions $\psi_{j,k}$ for $1\leq j \leq 2N$, $1\leq k
\leq 2N$ in the same way, the term $X_{k}$ is only replaced by
$Y_{k}$.
\begin{eqnarray}
\label{fonc3}
\begin{array}{l} \psi_{4j+1,k} = \gamma_{k}^{4j-1} \sin
Y_{k}, \quad \psi_{4j+2,k} =
\gamma_{k}^{4j} \cos Y_{k},  \\
\psi_{4j+3,k} = - \gamma_{k}^{4j+1} \sin Y_{k}, \quad \psi_{4j+4,k}
= - \gamma_{k}^{4j+2} \cos Y_{k},
\end{array}
\end{eqnarray}
for $1 \leq k \leq N$, and
\begin{eqnarray}
\label{fonc4}
\begin{array}{l} \psi_{4j+1,N+k} = \gamma_{k}^{2N-4j-2}
\cos Y_{N+k}, \quad \psi_{4j+2,N+k} =-
\gamma_{k}^{2N-4j-3} \sin Y_{N+k},  \\
\psi_{4j+3,N+k} = - \gamma_{k}^{2N-4j-4} \cos Y_{N+k}, \quad
\psi_{4j+4,N+k} = \gamma_{k}^{2N-4j-5} \sin Y_{N+k},
\end{array}
\end{eqnarray}
for $1  \leq k \leq N$.\\
Then it is clear that
$$
q(x,t):=\frac { W_{3}  (0) } { W_{1}  (0) }
$$
can be written as
\begin{eqnarray}
\label{qmod} q(x,t) = \frac { \Delta_{3}  } { \Delta_{1} }  = \frac
{ \det (\varphi_{j,k})_{j,\,k \in [1,2N]}  } {   \det
(\psi_{j,k})_{j,\,k \in [1,2N]} }.
\end{eqnarray}
We recall that $\lambda_{j} =1- 2j \epsilon^{2}$. All the functions
$\varphi_{j,k}$ and $\psi_{j,k}$ and their derivatives depend on
$\epsilon$ and can all be
prolonged by continuity when $\epsilon=0$.\\
Then the following expansions are used
$$
\varphi_{j,k}(x,t,\epsilon) = \sum_{l=0}^{N-1} \frac{1}{(2l)!}
\varphi_{j,1}[l] k^{2l} \epsilon^{2l} + O(\epsilon^{2N}), \quad
\varphi_{j,1}[l] = \frac{\partial^{2l} \varphi_{j,1}}
{\partial\epsilon^{2l}}(x,t,0),
$$
$$
\varphi_{j,1}[0] =  \varphi_{j,1} (x,t,0),  \quad 1 \leq j \leq 2N,
\quad 1 \leq k \leq N, \quad 1 \leq l \leq N-1,
$$
$$
\varphi_{j,N+k}(x,t,\epsilon) = \sum_{l=0}^{N-1} \frac{1}{(2l)!}
\varphi_{j,N+1}[l] k^{2l} \epsilon^{2l} + O(\epsilon^{2N}), \quad
\varphi_{j,N+1}[l] = \frac{\partial^{2l} \varphi_{j,N+1}}
{\partial\epsilon^{2l}}(x,t,0) ,
$$
$$
\varphi_{j,N+1}[0] =  \varphi_{j,N+1} (x,t,0),  \quad 1 \leq j \leq
2N, \quad 1 \leq k \leq N, \quad 1 \leq l \leq N-1.
$$
We have the same expansions for the functions $\psi_{j,k}$.
$$
\psi_{j,k}(x,t,\epsilon) = \sum_{l=0}^{N-1} \frac{1}{(2l)!}
\psi_{j,1}[l] k^{2l} \epsilon^{2l} + O(\epsilon^{2N}), \quad
\psi_{j,1}[l] = \frac{\partial^{2l} \psi_{j,1}}
{\partial\epsilon^{2l}}(x,t,0),
$$
$$
\psi_{j,1}[0] =  \psi_{j,1} (x,t,0),  \quad 1 \leq j \leq 2N, \quad
1 \leq k \leq N, \quad 1 \leq l \leq N-1,
$$

$$
\psi_{j,N+k}(x,t,\epsilon) = \sum_{l=0}^{N-1} \frac{1}{(2l)!}
\psi_{j,N+1}[l] k^{2l} \epsilon^{2l} + O(\epsilon^{2N}), \quad
\psi_{j,N+1}[l] =
 \frac{\partial^{2l} \psi_{j,N+1}}
{\partial\epsilon^{2l}}(x,t,0) ,
$$
$$
\psi_{j,N+1}[0] =  \psi_{j,N+1} (x,t,0),  \quad 1 \leq j \leq 2N,
\quad 1 \leq k \leq N, \quad N+1 \leq k \leq 2N..
$$
Then we get the following result :
\begin{thar}
The function $v$ defined by
\begin{eqnarray}
\label{soldet} v(x,t) = \exp (2it -i \varphi)\times
\frac{\det((n_{jk)_{j,k\in [1,2N]}})}{\det((d_{jk)_{j,k\in
[1,2N]}})}
\end{eqnarray}
is a quasi-rational solution of the NLS equation (\ref{NLS1})
$$
i v_{t}+ v_{xx} + 2|v|^{2}v=0,
$$
where
\begin{eqnarray}
\begin{array}{l}
\label{detquo} n_{j1}=\varphi_{j,1}(x,t,0), \, 1\leq j \leq 2N \quad
n_{jk}= \frac{\partial^{2k-2} \varphi_{j,1}}
{\partial\epsilon^{2k-2}}(x,t,0), \, \nonumber
\\
n_{jN+1}=\varphi_{j,N+1}(x,t,0), \, 1\leq j \leq 2N \quad n_{jN+k}=
\frac{\partial^{2k-2} \varphi_{j,N+1}}
{\partial\epsilon^{2k-2}}(x,t,0), \, \nonumber
\\
d_{j1}=\psi_{j,1}(x,t,0), \, 1\leq j \leq 2N \quad d_{jk}=
\frac{\partial^{2k-2} \psi_{j,1}} {\partial\epsilon^{2k-2}}(x,t,0),
 \nonumber
\\d_{jN+1}=\psi_{j,N+1}(x,t,0), \, 1\leq j \leq 2N \quad d_{jN+k}=
\frac{\partial^{2k-2} \psi_{j,N+1}}
{\partial\epsilon^{2k-2}}(x,t,0), \, \nonumber \\
2\leq k \leq N , \, 1\leq j \leq 2N
\end{array}
\end{eqnarray}
The functions $\varphi$ and $\psi$ are defined in
(\ref{fonc1}),(\ref{fonc2}), (\ref{fonc3}), (\ref{fonc4}).
\end{thar}
\noindent \textbf{Proof : } The columns of the determinants
appearing in $q(x,t)$  are combined successively to eliminate in
each column $k$ (and $N+k$) of them the powers of $\epsilon$
strictly inferior to $2(k-1)$; then each common term in numerator and denominator is
factorized and simplified; finally we take the limit when $\epsilon$ goes to $0$. \\
Precisely, first of all, the components $j$ of the columns $1$ and
$N+1$ are respectively equal by definition to $\varphi_{j1}[0]+
0(\epsilon^{})$ for $C_{1}$, $\varphi_{jN+1}[0]+ 0(\epsilon^{})$ for
$C_{N+1}$ of $\Delta_{3}$, and $\psi_{j1}[0]+ 0(\epsilon^{})$ for
$C'_{1}$, $\psi_{jN+1}[0]+ 0(\epsilon^{})$ for $C'_{N+1}$ of
$\Delta_{1}$.\\
At the first step of the reduction, we replace the columns $C_{k}$
by $C_{k}-C_{1}$ and $C_{N+k}$ by $C_{N+k}-C_{N+1}$ for $2\leq k
\leq N$, for $\Delta_{3}$; the same changes for $\Delta_{1}$ are
done. Each component $j$ of the column $C_{k}$ of $\Delta_{3}$ can
be rewritten as
$\sum_{l=1}^{N-1}\frac{1}{(2l)!}\varphi_{j,1}[l](k^{2l}-1)\epsilon^{2l}$
and the column $C_{N+k}$ replaced by
$\sum_{l=1}^{N-1}\frac{1}{(2l)!}\varphi_{j,N+1}[l](k^{2l}-1)\epsilon^{2l}$
for $2\leq k \leq N$. For $\Delta_{1}$, we have the same reductions,
each component $j$ of the column $C'_{k}$ can be rewritten as
$\sum_{l=1}^{N-1}\frac{1}{(2l)!}\psi_{j,1}[l](k^{2l}-1)\epsilon^{2l}$
and the column $C'_{N+k}$ replaced by
$\sum_{l=1}^{N-1}\frac{1}{(2l)!}\psi_{j,N+1}[l](k^{2l}-1)\epsilon^{2l}$
for $2\leq k \leq N$. \\
The term $\frac{k^{2}-1}{2}\epsilon^{2}$ for $2 \leq k \leq N$ can
factorized in $\Delta_{3}$ and $\Delta_{1}$ in each column $k$ and
$N+k$ , and so these common terms can be simplified in numerator and
denominator.\\
If we restrict the developments at order $1$ in columns $2$ and
$N+2$, we get respectively $\varphi_{j1}[1]+ 0(\epsilon^{})$ for
component $j$ of $C_{2}$, $\varphi_{jN+1}[1]+ 0(\epsilon^{})$ for
component $j$ of $C_{N+2}$ of $\Delta_{3}$, and $\psi_{j1}[1]+
0(\epsilon^{})$ for component $j$ of $C'_{2}$, $\psi_{jN+1}[1]+
0(\epsilon^{})$ for component $j$ of $C'_{N+2}$ of $\Delta_{1}$.
This algorithm can be continued up to the columns $C_{N}$, $C_{2N}$
of $\Delta_{3}$ and $C'_{N}$,
$C'_{2N}$ of $\Delta_{1}$. \\
Then taking the limit when $\epsilon$ tends to $0$, $q(x,t)$ can be
replaced by $Q(x,t)$ defined by :
\begin{eqnarray}
\label{Q} Q(x,t):= \frac {\left|
\begin{array}{cccccc}
\varphi_{1,1}[0] &  \ldots & \varphi_{1,1}[N-1] & \varphi_{1,N+1}[0] & \ldots & \varphi_{1,N+1}[N-1]\\
\varphi_{2,1}[0] &  \ldots & \varphi_{2,1}[N-1] & \varphi_{2,N+1}[0] & \ldots & \varphi_{2,N+1}[N-1]\\
\vdots  & \vdots & \vdots & \vdots  & \vdots & \vdots  \\
\varphi_{2N,1}[0] &  \ldots & \varphi_{2N,1}[N-1] & \varphi_{2N,N+1}[0] & \ldots & \varphi_{2N,N+1}[N-1]\\
\end{array}
\right|} {\left|
\begin{array}{cccccc}
\psi_{1,1}[0] &  \ldots & \psi_{1,1}[N-1] & \psi_{1,N+1}[0] & \ldots & \psi_{1,N+1}[N-1]\\
\psi_{2,1}[0] &  \ldots & \psi_{2,1}[N-1] & \psi_{2,N+1}[0] & \ldots & \psi_{2,N+1}[N-1]\\
\vdots  & \vdots & \vdots & \vdots  & \vdots & \vdots  \\
\psi_{2N,1}[0] &  \ldots & \psi_{2N,1}[N-1] & \psi_{2N,N+1}[0] & \ldots & \psi_{2N,N+1}[N-1]\\
\end{array}
\right|}
\end{eqnarray}
So the solution of the NLS equation takes the form :
$$
v(x,t) = \exp (2it -i \varphi)\times Q(x,t)
$$
So we get the result given in $(\ref{soldet})$. $\Box$\\


\section{Families of quasi-rational solutions of order $N$ depending on $2N-2$ parameters}
Here a theorem which states the structure of the quasi-rational
solutions to the NLS equation is given. It was only conjectured in
preceding works \cite{Gaillard10, Gaillard17}. Moreover we obtain
here families depending on $2N-2$ parameters for the $N$th-order
Peregrine breather including families with $2$ parameters
constructed in preceding works and so we get other symmetries
in these deformations than those were expected. \\
In this section we use the notations defined in the previous
sections. The functions $\varphi$ and $\psi$ are defined in
(\ref{fonc1}), (\ref{fonc2}), (\ref{fonc3}), (\ref{fonc4}).

\begin{thar}
The function $v$ defined by
\begin{eqnarray}
\label{soldetR} v(x,t) = \exp (2it -i \varphi)\times
\frac{\det((n_{jk)_{j,k\in [1,2N]}})}{\det((d_{jk)_{j,k\in
[1,2N]}})}
\end{eqnarray}
is a quasi-rational solution of the NLS equation (\ref{NLS1})
quotient of two polynomials $N(x,t)$ and $D(x,t)$ depending on
$2N-2$ real parameters
$\tilde{a_{j}}$ and $\tilde{b_{j}}$, $1 \leq j \leq N-1$.\\
$N$ and $D$ are polynomials of degrees $N(N+1)$ in $x$ and $t$.\\
\end{thar}
\textbf{Proof : } From the previous result (\ref{Q}), we need to
analyze functions $\varphi_{k,1}$, $\psi_{k,1}$ and
$\varphi_{k,N+1}$, $\psi_{k,N+1}$. Functions $\varphi_{k,j}$ and
$\psi_{k,j}$ differ only by the term of the argument $x_{3,k}$, so
only the study of functions $\varphi_{k,j}$ will be carried out.
Then the study of functions $\psi_{k,j}$ can be easily deduced from the analysis of $\varphi_{k,j}$.\\
The expansions of these functions in $\epsilon$ are studied. We
denote $(l_{kj})_{k,j \in [1,2N]}$ the matrix defined by
$$
l_{kj} = \frac{\partial^{2j-2}}{\partial \epsilon^{2j-2}}
\varphi_{k1}, \quad l_{k,j+N} = \frac{\partial^{2j-2}}{\partial
\epsilon^{2j-2}} \varphi_{k,1+N}, \quad 1\leq j \leq N, \,  1\leq k
\leq 2N,
$$
$\frac{\partial^{0}}{\partial x^{0}} \varphi_{}$ meaning $\varphi$.
Each coefficient of the matrix $(l_{kj})_{k,j \in [1,2N]}$ must be
evaluated, the power of $x$ and $t$ in the coefficient of
$\epsilon^{2(m-1)}$ for the column $m \in [1,2N]$. We remark that
with these notations, the matrix $(l_{kj})_{k,j \in [1,2N]}$
evaluated in $\epsilon = 0 $ is exactly $(n_{kj})_{k,j \in [1,2N]}$
defined in (\ref{detquo}).
Four cases must be studied depending on the parity of $k$.\\
\noindent 1. We study $l_{k1}$ for $k$ odd, $k=2s+1$. \\
$$
 l_{k1} =(-1)^{s}\sin (2 \epsilon(1- \epsilon^{2})^{\frac{1}{2} }x+4 i
\epsilon(1- \epsilon^{2})^{\frac{1}{2}}(1-2 \epsilon^{2})t
$$
$$
- i \ln \frac{1+i \epsilon(1- \epsilon^{2})^{-\frac{1}{2}}} {1-i
\epsilon(1- \epsilon^{2})^{-\frac{1}{2}}} - e_{1}) \times
\epsilon^{k-2}(1- \epsilon^{2})^{-\frac{k-2}{2}}
$$
$$
=(-1)^{s}\sin  \epsilon( \sum_{l=0}^{p} c_{2l} \epsilon^{2l}x +2i
\sum_{l=0}^{p}c_{2l} \epsilon^{2l}(1-2 \epsilon^{2})t + 2
\sum_{l=0}^{p} (-1)^{l} \epsilon^{2l} \frac{(1-
\epsilon^{2})^{-\frac{2l+1}{2}}}{(2l+1)}
$$
$$
-\sum_{l=1}^{N-1} \tilde{a}_{l} \epsilon^{2l} + i \sum_{l=1}^{N-1}
\tilde{b}_{l} \epsilon^{2l} + O(\epsilon^{p+1}) ) \times
\epsilon^{k-2} (\sum_{l=1}^{r} g_{2l}\epsilon^{2l} +
O(\epsilon^{r+1}) )
$$
$$
= (-1)^{s}\sin  \epsilon( \sum_{l=0}^{p} (c_{2l} x + d_{2l} t +
f_{2l} + O(\epsilon^{p+1}) ) \epsilon^{2l} ) \times \epsilon^{k-2}
(\sum_{l=1}^{r} g_{2l}\epsilon^{2l} + O(\epsilon^{r+1}) )
$$
$$
= \sum_{l=0}^{q} \frac{(-1)^{l+s} \epsilon^{2l}}{(2l+1)!} (
\sum_{n=0}^{p} (c_{2n} x + d_{2n} t + f_{2n} + O(\epsilon^{p+1}) )
\epsilon^{2n})^{2l+1}
 \times \epsilon^{k-1} (\sum_{l=1}^{r}
g_{2l}\epsilon^{2l} + O(\epsilon^{r+1}) )
$$
$$
= \sum_{l=0}^{q} \frac{(-1)^{l+s} \epsilon^{2l}}{(2l+1)!} (
\sum_{n=0}^{p} P_{n} (x,t)  \epsilon^{2n})^{2l+1} \times
\epsilon^{k-1} \sum_{l=1}^{r} g_{2l}\epsilon^{2l} + O(\epsilon^{t})
$$
where $P_{n}(x,t)$ is a polynomial of order $1$ in $x$ and $t$.
$$
l_{k,1}= \sum_{l=0}^{q}  \epsilon^{2l} \sum_{\alpha_{0}+ \ldots +
\alpha_{p} = 2l+1} \beta_{\alpha_{0},\ldots,\alpha_{p}}
P_{0}(x,t)^{\alpha_{0}} \ldots P_{p}(x,t)^{\alpha_{p}}
 \epsilon^{ 2(\alpha_{1}+2\alpha_{2}+ p \alpha_{p}) } \times \epsilon^{2s} \sum_{l=1}^{r}
g_{2l}\epsilon^{2l} + O(\epsilon^{t})
$$
$$
= \sum_{l=0}^{q}  \epsilon^{2l} \sum_{\alpha_{0}+ \ldots +
\alpha_{p} = 2l+1} Q_{\alpha_{0},\dots, \alpha_{p}}(x,t)
 \epsilon^{ 2(\alpha_{1}+2\alpha_{2}+ p \alpha_{p}) } \times \epsilon^{2s} \sum_{l=1}^{r}
g_{2l} \epsilon^{2l} + O(\epsilon^{t}) ,
$$
where $Q_{\alpha_{0},\dots, \alpha_{p}}(x,t)$ is a polynomial of
order $2l+1$ in $x$ and $t$. \\
The terms in $ \epsilon^{0}$ are obtained for $l=0$ in the two
summations with $\alpha_{0}=1$.\\
For column $m$, we search the terms in $ \epsilon^{2m-2}$ with the
maximal power in $x$ and $t$. It is obtained for $2l+k-1 =
2m-2$, which gives $l=m-s-1$. \\
The notations given in (\ref{soldet}) are used. We get the following
result
\begin{proar}
\label{deg1}
\begin{eqnarray} \deg(n_{2s+1,m}) = 2(m-s)-1 \mbox{ for
} s\leq m-1, \quad n_{2s+1,m}=0 \mbox{ for } s \geq m.
\end{eqnarray}
\end{proar}
\noindent 2. We study $l_{k1}$ for $k$ even, $k=2s$. \\
$$
 l_{k1} = (-1)^{s+1} \cos (2 \epsilon(1- \epsilon^{2})^{\frac{1}{2} }x+4 i
\epsilon(1- \epsilon^{2})^{\frac{1}{2}}(1-2 \epsilon^{2})t
$$
$$
- i \ln \frac{1+i \epsilon(1- \epsilon^{2})^{-\frac{1}{2}}} {1-i
\epsilon(1- \epsilon^{2})^{-\frac{1}{2}}} - e_{1}) \times
\epsilon^{k-2}(1- \epsilon^{2})^{-\frac{k-2}{2}}
$$
$$
=(-1)^{s+1} \cos  \epsilon( \sum_{l=0}^{p} c_{2l} \epsilon^{2l}x +2i
\sum_{l=0}^{p}c_{2l} \epsilon^{2l}(1-2 \epsilon^{2})t + 2
\sum_{l=0}^{p} (-1)^{l} \epsilon^{2l} \frac{(1-
\epsilon^{2})^{-\frac{2l+1}{2}}}{(2l+1)}
$$
$$
-\sum_{l=1}^{N-1} \tilde{a}_{l} \epsilon^{2l} + i \sum_{l=1}^{N-1}
\tilde{b}_{l} \epsilon^{2l} + O(\epsilon^{p+1}) ) \times
\epsilon^{k-2} (\sum_{l=1}^{r} g_{2l}\epsilon^{2l} +
O(\epsilon^{r+1}) )
$$
$$
= (-1)^{s+1} \cos  \epsilon( \sum_{l=0}^{p} (c_{2l} x + d_{2l} t +
f_{2l} + O(\epsilon^{p+1}) ) \epsilon^{2l} ) \times \epsilon^{k-2}
(\sum_{l=1}^{r} g_{2l}\epsilon^{2l} + O(\epsilon^{r+1}))
$$
$$
= \sum_{l=0}^{q} \frac{(-1)^{l+d+1} \epsilon^{2l}}{(2l)!} (
\sum_{n=0}^{p} (c_{2n} x + d_{2n} t + f_{2n} + O(\epsilon^{p+1}) )
\epsilon^{2n})^{2l}
 \times \epsilon^{k-2} (\sum_{l=1}^{r}
g_{2l}\epsilon^{2l} + O(\epsilon^{r+1}) )
$$
$$
= \sum_{l=0}^{q} \frac{(-1)^{l+s+1} \epsilon^{2l}}{(2l)!} (
\sum_{n=0}^{p} P_{n} (x,t)  \epsilon^{2n})^{2l} \times
\epsilon^{2s-2} \sum_{l=1}^{r} g_{2l}\epsilon^{2l} + O(\epsilon^{t})
$$
where $P_{n}(x,t)$ is a polynomial of order $1$ in $x$ and $t$.
$$
 l_{k,1}= \sum_{l=0}^{q}  \epsilon^{2l} \sum_{\alpha_{0}+ \ldots +
\alpha_{p} = 2l} \beta_{\alpha_{0},\ldots,\alpha_{p}}
P_{0}(x,t)^{\alpha_{0}} \ldots P_{p}(x,t)^{\alpha_{p}}
 \epsilon^{ 2(\alpha_{1}+2\alpha_{2}+ p \alpha_{p}) } \times \epsilon^{2s-2} \sum_{l=1}^{r}
g_{2l}\epsilon^{2l} + O(\epsilon^{t})
$$
$$
= \sum_{l=0}^{q}  \epsilon^{2l} \sum_{\alpha_{0}+ \ldots +
\alpha_{p} = 2l} Q_{\alpha_{0},\dots, \alpha_{p}}(x,t)
 \epsilon^{ 2(\alpha_{1}+2\alpha_{2}+ p \alpha_{p}) } \times \epsilon^{2s-2} \sum_{l=1}^{r}
g_{2l} \epsilon^{2l} + O(\epsilon^{t}) ,
$$
where $Q_{\alpha_{0},\dots, \alpha_{p}}(x,t)$ is a polynomial of
order $2l$ in $x$ and $t$. \\
The terms in $ \epsilon^{0}$ are obtained for $l=0$ in the two
summations with $\alpha_{0}=1$.\\
For column $m$, we search the terms in $ \epsilon^{2m-2}$ with the
maximal power in $x$ and $t$. It is obtained for $2l+k-2 =
2m-2$, which gives $l=m-s$. \\
With the notations given in (\ref{soldet}), we have
\begin{proar}
\label{deg2}
\begin{eqnarray} \deg(n_{2s,m}) = 2(m-s) \mbox{ for
} s\leq m, \quad n_{2s,m}=0 \mbox{ for } s > m.
\end{eqnarray}
\end{proar}
\noindent 3. We study $l_{k \frac{M}{2}+1}$ for $k$ odd, $k=2s+1$. \\
$$
 l_{k \frac{M}{2}+1} = (-1)^{s} \cos (2 \epsilon(1-
\epsilon^{2})^{\frac{1}{2} }x - 4 i \epsilon(1-
\epsilon^{2})^{\frac{1}{2}}(1-2 \epsilon^{2})t + i \ln \frac{1+i
\epsilon(1- \epsilon^{2})^{-\frac{1}{2}}} {1-i \epsilon(1-
\epsilon^{2})^{-\frac{1}{2}}} - e_{\frac{M}{2}+1})
$$
$$
\times \epsilon^{M-k-1}(1- \epsilon^{2})^{-\frac{M-k-1}{2}}
$$
$$
=(-1)^{s} (\cos  \epsilon( \sum_{l=0}^{p} c_{2l} \epsilon^{2l}x - 2i
\sum_{l=0}^{p}c_{2l} \epsilon^{2l}(1-2 \epsilon^{2})t - 2
\sum_{l=0}^{p} (-1)^{l} \epsilon^{2l} \frac{(1-
\epsilon^{2})^{-\frac{2l+1}{2}}}{(2l+1)}
$$
$$
-\sum_{l=1}^{N-1} \tilde{a}_{l} \epsilon^{2l} + i \sum_{l=1}^{N-1}
\tilde{b}_{l} \epsilon^{2l} + O(\epsilon^{p+1}) ) \times
\epsilon^{M-k-1} (\sum_{l=1}^{r} g_{2l}\epsilon^{2l} +
O(\epsilon^{r+1}) )
$$
$$
=(-1)^{s} (\cos  \epsilon( \sum_{l=0}^{p} (c_{2l} x + d_{2l} t +
f_{2l}) \epsilon^{2l}  + O(\epsilon^{p+1}) ) \times \epsilon^{M-k-1}
(\sum_{l=1}^{r} g_{2l}\epsilon^{2l} + O(\epsilon^{r+1}) )
$$
$$
= \sum_{l=0}^{q} \frac{(-1)^{l+s} \epsilon^{2l}}{(2l)!} (
\sum_{n=0}^{p} (c_{2n} x + d_{2n} t + f_{2n} + O(\epsilon^{p+1}) )
\epsilon^{2n})^{2l}
 \times \epsilon^{M-k-1} (\sum_{l=1}^{r}
g_{2l}\epsilon^{2l} + O(\epsilon^{r+1}) )
$$
$$
= \sum_{l=0}^{q} \frac{(-1)^{l+s} \epsilon^{2l}}{(2l)!} (
\sum_{n=0}^{p} P_{n} (x,t)  \epsilon^{2n} + O(\epsilon^{p+1}) )^{2l}
\times \epsilon^{M-2s-2} (\sum_{l=1}^{r} g_{2l}\epsilon^{2l} +
O(\epsilon^{r+1}) )
$$
where $P_{n}(x,t)$ is a polynomial of order $1$ in $x$ and $t$.
$$
l_{k,\frac{M}{2}+1}= \sum_{l=0}^{q}  \epsilon^{2l} \sum_{\alpha_{0}+
\ldots + \alpha_{p} = 2l} \beta_{\alpha_{0},\ldots,\alpha_{p}}
P_{0}(x,t)^{\alpha_{0}}
$$
$$
\ldots P_{p}(x,t)^{\alpha_{p}}
 \epsilon^{ 2(\alpha_{1}+2\alpha_{2}+ p \alpha_{p}) } \times \epsilon^{M-2s-2} \sum_{l=1}^{r}
g_{2l}\epsilon^{2l} + O(\epsilon^{t})
$$
$$
= \sum_{l=0}^{q}  \epsilon^{2l} \sum_{\alpha_{0}+ \ldots +
\alpha_{p} = 2l} Q_{\alpha_{0},\dots, \alpha_{p}}(x,t)
 \epsilon^{ 2(\alpha_{1}+2\alpha_{2}+ p \alpha_{p}) } \times \epsilon^{M-2s-2} \sum_{l=1}^{r}
g_{2l} \epsilon^{2l} + O(\epsilon^{t}) ,
$$
where $Q_{\alpha_{0},\dots, \alpha_{p}}(x,t)$ is a polynomial of
order $2l$ in $x$ and $t$. \\
The terms in $ \epsilon^{0}$ (column $\frac{M}{2}+1$) are obtained
for $l=0$ in the two summations with $\alpha_{0}=1$.\\
For column $\frac{M}{2}+m$, we search the terms in $
\epsilon^{2m-2}$ with the maximal power in $x$ and $t$. It is
obtained for
$2l+2(N-s-1) = 2m-2$, which gives $l=m+s-N$. \\
Then we get the following result
\begin{proar}
\label{deg3}
\begin{eqnarray} \deg(n_{2s+1,m+\frac{M}{2}}) = 2m+2s-M \mbox{ for
} s \geq \frac{M}{2}-m, \quad n_{2s+1,m}=0 \mbox{ for } s <
\frac{M}{2} - m.
\end{eqnarray}
\end{proar}
\noindent 4. We study $l_{k,1+\frac{M}{2}}$ for $k$ even, $k=2s$. \\
$$
 l_{k \frac{M}{2}+1} = (-1)^{s} \sin (2 \epsilon(1-
\epsilon^{2})^{\frac{1}{2} }x-4 i  \epsilon(1-
\epsilon^{2})^{\frac{1}{2}}(1-2 \epsilon^{2})t + i \ln \frac{1+i
\epsilon(1- \epsilon^{2})^{-\frac{1}{2}}} {1-i \epsilon(1-
\epsilon^{2})^{-\frac{1}{2}}} - e_{\frac{M}{2}+1})
$$
$$
\times \epsilon^{M-k-1}(1- \epsilon^{2})^{-\frac{M-k-1}{2}}
$$
$$
=(-1)^{s} \sin  \epsilon( \sum_{l=0}^{p} c_{2l} \epsilon^{2l}x -2i
\sum_{l=0}^{p}c_{2l} \epsilon^{2l}(1-2 \epsilon^{2})t - 2
\sum_{l=0}^{p} (-1)^{l} \epsilon^{2l} \frac{(1-
\epsilon^{2})^{-\frac{2l+1}{2}}}{(2l+1)}
$$
$$
-\sum_{l=1}^{N-1} \tilde{a}_{l} \epsilon^{2l} + i \sum_{l=1}^{N-1}
\tilde{b}_{l} \epsilon^{2l} + O(\epsilon^{p+1}) ) \times
\epsilon^{M-k-1} (\sum_{l=1}^{r} g_{2l}\epsilon^{2l} +
O(\epsilon^{r+1}) )
$$
$$
= (-1)^{s} \sin  \epsilon( \sum_{l=0}^{p} (c_{2l} x + d_{2l} t +
f_{2l}) \epsilon^{2l}  + O(\epsilon^{p+1}) ) \times \epsilon^{M-k-1}
(\sum_{l=1}^{r} g_{2l}\epsilon^{2l} + O(\epsilon^{r+1}) )
$$
$$
= \sum_{l=0}^{q} \frac{(-1)^{l+s} \epsilon^{2l}}{(2l+1)!} (
\sum_{n=0}^{p} (c_{2n} x + d_{2n} t + f_{2n} + O(\epsilon^{p+1}) )
\epsilon^{2n})^{2l+1}
 \times \epsilon^{M-k} (\sum_{l=1}^{r}
g_{2l}\epsilon^{2l} + O(\epsilon^{r+1}) )
$$
$$
= \sum_{l=0}^{q} \frac{(-1)^{l+s} \epsilon^{2l}}{(2l+1)!} (
\sum_{n=0}^{p} P_{n} (x,t)  \epsilon^{2n} + O(\epsilon^{p+1})
)^{2l+1} \times \epsilon^{M-2s} (\sum_{l=1}^{r} g_{2l}\epsilon^{2l}
+ O(\epsilon^{r+1}) )
$$
where $P_{n}(x,t)$ is a polynomial of order $1$ in $x$ and $t$.
$$
l_{k,1}= \sum_{l=0}^{q}  \epsilon^{2l} \sum_{\alpha_{0}+ \ldots +
\alpha_{p} = 2l+1} \beta_{\alpha_{0},\ldots,\alpha_{p}}
P_{0}(x,t)^{\alpha_{0}}
$$
$$
\ldots P_{p}(x,t)^{\alpha_{p}}
 \epsilon^{ 2(\alpha_{1}+2\alpha_{2}+ p \alpha_{p}) } \times \epsilon^{M-2s} \sum_{l=1}^{r}
g_{2l}\epsilon^{2l} + O(\epsilon^{t})
$$
$$
= \sum_{l=0}^{q}  \epsilon^{2l} \sum_{\alpha_{0}+ \ldots +
\alpha_{p} = 2l+1} Q_{\alpha_{0},\dots, \alpha_{p}}(x,t)
 \epsilon^{ 2(\alpha_{1}+2\alpha_{2}+ p \alpha_{p}) } \times \epsilon^{M-2s} \sum_{l=1}^{r}
g_{2l} \epsilon^{2l}  + O(\epsilon^{t}) ,
$$
where $Q_{\alpha_{0},\dots, \alpha_{p}}(x,t)$ is a polynomial of
order $2l+1$ in $x$ and $t$. \\
The terms in $ \epsilon^{0}$ are obtained for $l=0$ in the two
summations with $\alpha_{0}=1$.\\
For column $\frac{M}{2}+m$, we search the terms in $
\epsilon^{2m-2}$ with the maximal power in $x$ and $t$. It is
obtained for $2l+M-k =
2m-2$, which gives $l=m+s-N-1$. \\
Using the notations given in (\ref{soldet}), we get the following
result
\begin{proar}
\label{deg4}
\begin{eqnarray}
\begin{array}{l}
\deg(n_{2s,m+\frac{M}{2}}) = 2m + 2s - M - 1 \mbox{ for
} s\geq \frac{M}{2}+1-M, \quad  \\
n_{2s,m+\frac{M}{2}}=0 \mbox{ for } s < \frac{M}{2}+1-m.
\end{array}
\end{eqnarray}
\end{proar}
These results can be rewritten in the following way
\begin{proar}
\begin{eqnarray}
\label{degtot}
\begin{array}{l}
\deg(n_{j,k}) = 2k - j  \mbox{ for } j \leq 2k, \quad \\
n_{j,k} = 0 \mbox{ for } j > 2k, \\
\deg(n_{j,k}) = 2k + j -2M - 1 \mbox{ for } j \geq 2 M + 1 - 2k,
\quad \\
n_{j,k} = 0 \mbox{ for } j < 2 M + 1 - 2k.
\end{array}
\end{eqnarray}
\end{proar}
The degree of the determinant of the matrix
$(n_{kj})_{k,j \in [1,2N]}$ can now be evaluated. \\
From the previous analysis, we see that $x$ and $t$ have necessarily
the same power in each $n_{kj}$. The maximal power in $x$ and $t$,
is successively taken in each column. It is realized by the
following product
$$
\prod_{j=1}^{N} n_{j,j} \prod_{j=1}^{N} n_{N+j,2N+1-j}.
$$
Applying the result given in (\ref{degtot}) we get
$$
\deg ( \det (n_{kj})_{k,j \in [1,2N]} ) = \sum_{j=1}^{N} \deg
(n_{j,j}) + \sum_{j=1}^{N} \deg ( n_{N+j,2N+1-j})
$$
$$
= \sum_{j=1}^{N} 2j-j + \sum_{j=1}^{N} 2(M+1-j) - 2M-1 + \frac{M}{2}
+ j
$$
$$
= \sum_{j=1}^{N} j + \sum_{j=1}^{N} N+1-j = N(N+1).
$$
It is the same for determinant $\det (d_{kj})_{k,j \in [1,2N]}$,
we have $\deg ( \det (d_{kj})_{k,j \in [1,2N]} ) = N(N+1)$.\\
Thus the quotient
$$
\frac{\det((n_{kj)_{j,k\in [1,2N]}})}{\det((d_{kj)_{j,k\in
[1,2N]}})}
$$
defines a quotient of two polynomials, each of them of degree $N(N+1)$, and this proves the result. \\
Parameters $a_{1} = \sum_{k=1}^{N-1}\tilde{a}_{k}\epsilon_{k} $ and
$a_{1} = \sum_{k=1}^{N-1}\tilde{a}_{k}\epsilon_{k} $ must be
chosen in the following way. \\
The term $\epsilon_{k}$ must be a power of $\epsilon$ to get a
nontrivial solution; $\epsilon_{k}$ must be a strictly positive
number $a$ in order to have a finite limit when $\epsilon$ goes to
$0$. If the power of $\epsilon$ is superior to $2N-2$, the
derivations going up to $2N-2$, then this coefficient becomes $0$
when the limit is taken when $\epsilon$ goes to $0$ and so has no
relevance in the expression of the limit. \\
$\Box$


\section{Conclusion}
Here we proved the structure of quasi-rational solutions to the
one dimensional focusing NLS equation at order $N$. They can be
expressed as a product of an exponential depending on $t$ by a ratio
of two polynomials of degree $N(N+1)$ in $x$ and $t$.
If we choose $\tilde{a}_{i}=\tilde{b}_{i}=0$ for $1\leq i \leq N-1$,
we obtain the classical (analogue) Peregrine breather. Thus these
solutions appear as $2N-2$-parameters deformations of the Peregrine
breather of order $N$. \\
The solutions for orders $3$ and $4$ first found by Matveev have also been explicitly found
by the present author \cite{Gaillard27,Gaillard26}.
We have also explicitly found the solutions at order $5$ with $8$
parameters \cite{Gaillard29}: these expressions are too extensive to be presented :
it takes $14049$ pages! For other orders $6,\,7,\,8$, the solutions are also
explicitly found but are too long to be published in
any review. In the relative works  \cite{Gaillard28, Gaillard30, Gaillard36, Gaillard37, Gaillard33} only the analysis has been done and figures of deformations of the Peregrine breathers has been realized.
The solutions for order $9$ with $16$ parameters \cite{Gaillard37} and respectively for order $10$ with $18$ parameters are also completely found \cite{Gaillard33}. \\
We still insist on the fact that quasi rational solutions of NLS equation can be expressed as a quotient of two polynomials of degree $N(N+1)$ in $x$ and $t$ dependent on $2N-2$ real parameters by an exponential depending on time. Among these aforementioned solutions of order $N$, there is one which has the largest module : it is the solution obtained in this representation when all the parameters are equal to $0$; one obtains the Peregrine breather order $N$. His importance is due to the fact that among the solutions of order $N$, its module is largest, equal to $2N+1$. This result first formulated by Akhmediev has just been proved recently \cite{Gaillard39}. \\
In the recent studies proposed by the author, the solutions of order $N$ can be represented by their module in the plane $(x; t)$. With the representation given in this article, one obtains at order $N$, the configurations containing $N(N+1)/2$ peaks, except the special case of Peregrine breather. These configurations can be classified according to the values of the parameters $a_{i}$ or $b_{i}$ for $i$ varying between $1$ and $N-1$. It is important to note that the role played by $a_{i}$ or $b_{i}$ for a given index $i$ is the same one, in obtaining the configurations. The study refers to the analysis of the solutions when only one of the parameters is not equal to $0$. Among these solutions, one distinguishes two types of configurations; for $a_{1}$ or $b_{1}$ not equal to $0$, one observes triangular configurations with $N(N+1)/2$ peaks. For $a_{i}$ or $b_{i}$ not equal to $0$ and $2\leq i \leq N-1$, one observes concentric rings. The simplest structure is obtained for $a_ {N-1}$ or $b_ {N-1}$ not equal to $0$ : one obtains only one ring of $2N-1$ peaks with in his center Peregrine breather of order $N-2$; this fact was also first formulated by Akhmediev. The detailed study of the other structures is being analyzed. We hope to be able to give results soon. \\
We can conclude that the method described in the present paper provides a
very efficient and powerful tool to get explicit solutions to the NLS equation and to understand the behavior of rogue waves. \\
There are currently many applications in different fields as recent
works by Akhmediev et al. \cite{Chabchoub2} or Kibler et al.
\cite{Kibler} attest it in particular. \\
This study leads to a better understanding of the
phenomenon of rogue waves, and it would be relevant to go on with higher orders. \\



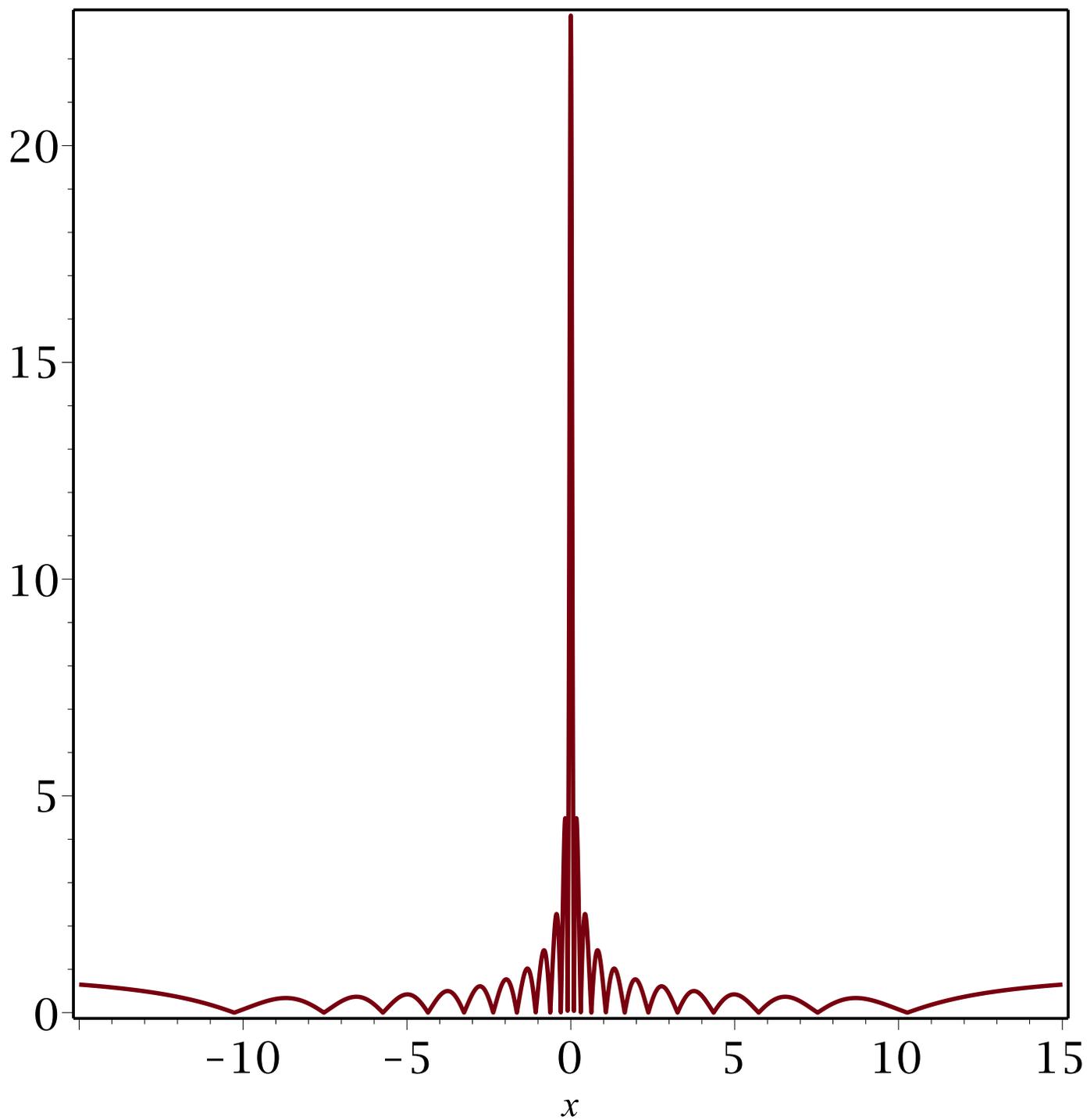
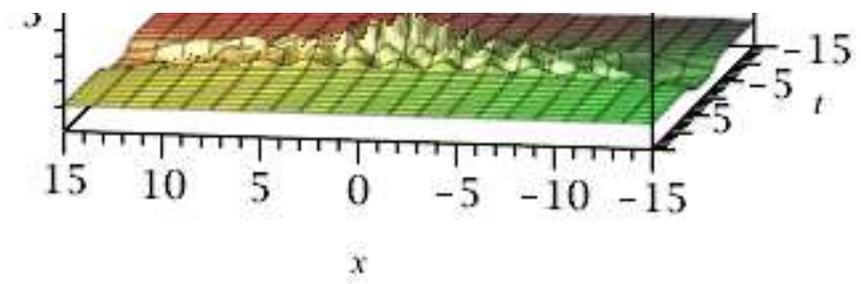
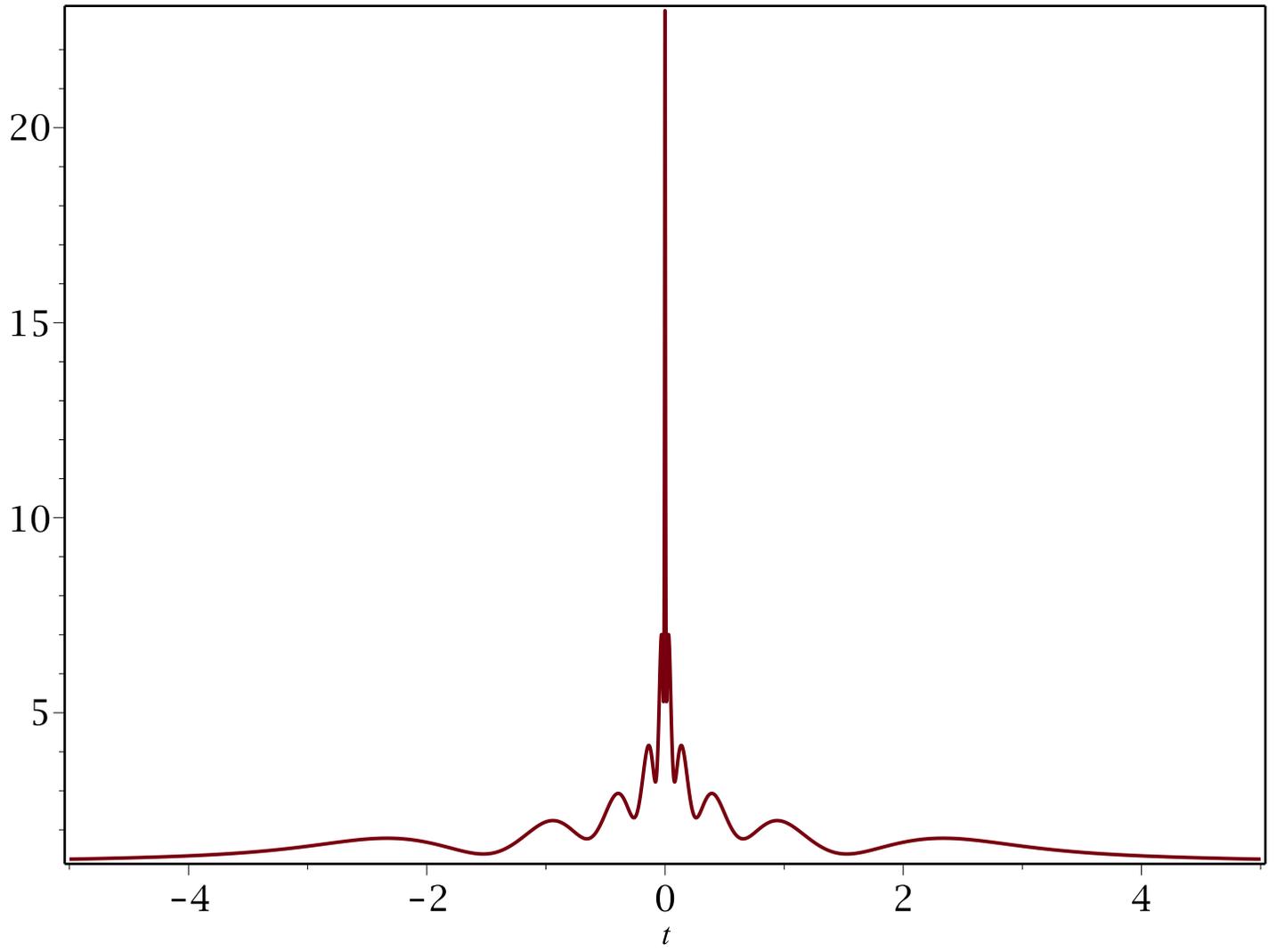
\end{document}